\documentclass[aps,APL,twocolumn,superscriptaddress,amsmath,amsfonts,
               floatfix]{revtex4}

\usepackage{epsfig,psfrag}
\usepackage{CJK}

\usepackage{graphicx}
\usepackage{dcolumn}
\usepackage{bm}
\usepackage{amssymb}

\begin{document}
\title{Multiphonons resonance Raman scattering in Landau-quantized graphene}
\author{Zi-Wu Wang}
\affiliation{Tianjin Key Laboratory of Low Dimensional Materials Physics and Preparing Technology,
Department of Physics, Tianjin University, Tianjin 300354 China}
\email{wangziwu@tju.edu.cn}
\author{Zhi-Qing Li}
\affiliation{Tianjin Key Laboratory of Low Dimensional Materials Physics and Preparing Technology,
Department of Physics, Tianjin University, Tianjin 300354 China}

\begin{abstract}
We theoretically investigate multiphonons resonance Raman scattering between the Landau levels in graphene on the polar substrate using the Huang-Rhys's model. We not only present the single and multiple surface optical (SO) phonons scattering, but also propose the combined multiphonons scattering (CMS), which is composed of the SO phonon and longitudinal acoustic phonon. We find that the CMS has a blue-shift behavior with increasing the magnetic field, differing from these SO phonon resonance scattering at a special magnetic field. This behavior may be used to explain the changing shoulder of the Raman spectrum of optical phonon resonance scattering in experiments. The theoretical model could be expanded to analyze the fine structure of Raman spectrum in two-dimensional materials.
\end{abstract}
\maketitle

The non-equidistant Landau level (LL) separation is an unique property of graphene when the perpendicular magnetic field is presented, which displays a number of interesting optical properties\cite{wl1,wl2}, such as the selection rule of the emission of polarized light\cite{wl3}, a giant Faraday rotation\cite{wl4}and generation of entangled photon states\cite{wl5}. In particular, the LL separation can be tuned from the infrared to terahertz regimes by manipulating the external magnetic field, which may be used as the infrared and terahertz sources for many electric devices\cite{wl6,wl7,wl8,wl9,wl10,wl11}. Therefore, the studies of the non-radiative relaxation channels between LLs, competing with the optical transitions processes, are of crucial for the exploration of optical properties in the Landau-quantized graphene (LQG).

Two types of non-radiation channels were mainly considered until now. One is the Auger scattering, which prevails at high carriers concentration and has been proved in experiments\cite{wl12,wl13}. Anther is the different optical phonons scattering, such as the longitudinal optical (LO) phonons\cite{wl3,wl14}, the surface optical (SO) phonons induced by the supporting substrates\cite{wl15} and the out-of-plane phonon at the $\Gamma$ point\cite{wl16}, which plays a dominant role when the LL's separation matching the optical phonon energy. This is known as the magnetophonon resonance effect, which has been identified widely in experiments by the magneto-exciton transition\cite{wl17}, magneto-optical conductivity\cite{wl18} and magneto-Raman scattering\cite{wl14,wl19}. However, theory predicted that this type of channel will be strongly suppressed as the optical phonon energy mismatches the LL's separation, which is also called the relaxation bottleneck effect. In order to break this effect, we have proposed the combined two-phonons scattering (CTPS)\cite{wl20,wl21} and find that its relaxation time is fast and very close to the Auger scattering. Auger scattering has been extensively proved in experiments, but the progress of the CTPS is very slow due to the lack of the proper platform and effective way to identify these CTPS.

In the present paper, we theoretically propose the Raman scattering with multiphonon processes in LQG based on the Huang-Rhys model. We give the single and multiple SO phonons as well as the CTPS composed of the SO and longitudinal acoustic (LA) phonon modes. We study the intensity of Raman scattering and present the modulations of the intensity of the scattering by the external magnetic filed, temperature, polarizability of substrate and internal distance between graphene and substrate. Our theoretical results indicate that the LQG provides a good platform to identify the multiphonons scattering, especially for the CTPS in experiments.
\begin{figure}
\includegraphics[width=3.0in,keepaspectratio]{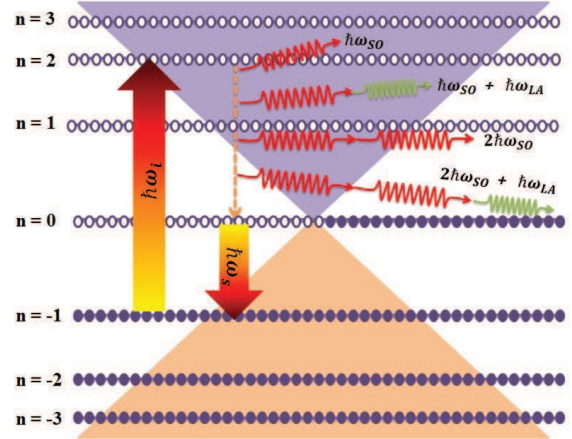}
\caption{\label{compare} The schematic diagrams for the distributions of the Landau levels in graphene and a possible channel of Raman scattering with different multiphonons resonance processes. $\hbar\omega_i$ and $\hbar\omega_s$ stand for the incident photon and scattering photon, respectively.}
\end{figure}
\begin{figure*}
\includegraphics[width=7.0in,keepaspectratio]{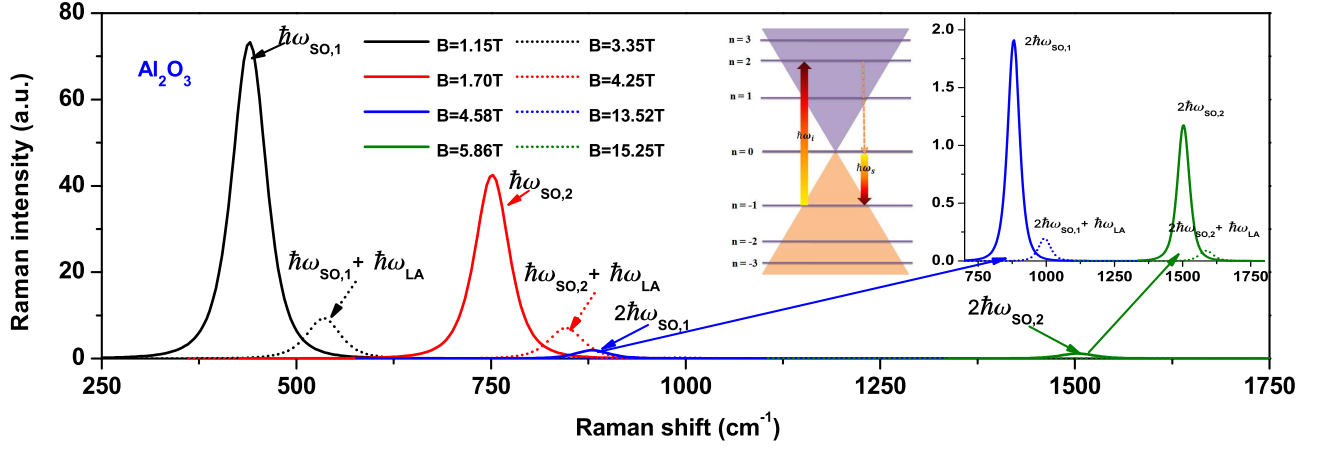}
\caption{Multiphonons resonance scattering in the processes of $-1\rightarrow2\rightsquigarrow0\rightarrow-1$ in Landau-quantized graphene on Al$_2$O$_3$ substrate at temperature $T=$ 80 K and the internal distance $z=$ 0.5 nm. The inset shows the intensity of two-SO phonon and combined three-phonons scattering.}
\end{figure*}

We consider that a uniform magnetic field $\textbf{B}$ is perpendicular to the monolayer graphene sheet laying on the polar substrate. In the frame of symmetric gauge for $\textbf{B}$, the energies are quantized into Landau levels schemed in Fig. 1 and
the corresponding eigenfunction and eigenenergy of the single free carrier can be solved analytically (see the supplementary material).

In general, Raman scattering involves electronic excitations as intermediate states. In LQG, the Landau level can be acted as the intermediate state for the resonance Raman scattering. The electromagnetic field of the incident and scattering photon interact primarily with these Landau levels, and emission of phonons occurs due to electron-phonon interaction. Based on the Huang-Rhys model\cite{wl22}, the cross section of multiphonons Raman scattering can be schematically represented as
\begin{widetext}
\begin{equation}
R_n\propto A_{n}\sum_{j,k,\nu}\mid\frac{\mu_{ij}(\omega_i)\mid\langle\psi_j|\langle\chi_j(Q_{j,\nu})|H_{ep}^{\nu}|\chi_k(Q_{k,\nu})\rangle|\psi_k\rangle\mid\mu_{ki}(\omega_s)}
{[\hbar\omega_i-(E_j-E_i)-i\Upsilon][\hbar\omega_i-\hbar\omega_s-\sum_{\nu}n\hbar\omega_{\nu}-i\Upsilon]}\mid^2,
\end{equation}
\end{widetext}
with
\begin{equation}
\mu_{ij}(\omega_i)=\frac{\hbar e^2V_F^2}{2\varepsilon_0\omega_i}\left\{\begin{aligned} \frac{1}{2}\quad (j=0 \qquad or \quad i=0)\\\frac{1}{4}\quad(j\neq0 \quad and \quad i\neq0)
\end{aligned}
\right.,
\end{equation}
which describes the transition process between LLs for the incident photon following the optical selection rule\cite{wl3,wl5}. The scattering process $\mu_{ki}(\omega_s)$ has the similar form. The detail approximation for $\mu_{ij}(\omega_i)$ and $\mu_{ki}(\omega_s)$ are in supplementary material. $\psi_j$ ($\psi_k$) and $E_j$ ($E_i$) is the eigenfunction and eigenvalue of LL, respectively. $\chi_j(Q_{j,\nu})$ ($\chi_k(Q_{k,\nu})$) is the harmonic oscillation describing the lattice vibration with the equilibrium position $Q_{j,\nu}$ ($Q_{k,\nu}$). In this paper, the linear electron-SO phonons and -LA phonons couplings $H_{ep}^{\nu}$($\nu$=SO,LA) are taken into account and can be given
\begin{equation}
H_{ep}^{\nu}=\sum_q M_{\nu}(q)e^{i q\cdot r}Q_{\nu,q}=\sum_q \mathcal{M}^{\nu}(q \cdot r)Q_{\nu,q},
\end{equation}
with the coupling element\cite{wl18,wl23}
\begin{equation}
M_{SO_{\lambda}}(q)=\sqrt{\frac{e^2 \hbar\omega_{SO,\lambda}}{2A\varepsilon_0}(\frac{1}{\kappa_0+1}-\frac{1}{\kappa_{\infty}+1})}e^{-qz}\nonumber\\,
\end{equation}
for SO phonon mode including two branches ($\lambda=1,2$) and
\begin{equation}
M_{LA}(q)=i\sqrt{\frac{\hbar D^2 q^2 }{A\rho\omega_{LA}}}\nonumber\\,
\end{equation}
for LA phonon mode in the deformation potential mechanism\cite{wl24}. The parameters in the two coupling elements are same as that of ref. (23) and (24). $n$ is the scattering phonon number and $\hbar\omega_{\nu}$ is the phonon energy with the mode $\nu$, $\Upsilon$ is the broadening factor. In the Franc-Condon approximation, the term of the phonon scattering in Eq. (1) can be rewritten as
\begin{eqnarray}
&&A_{n}\sum_{\nu}\mid\langle\psi_j|\langle\chi_j(Q_{j,\nu})|H_{ep}^{\nu}|\chi_k(Q_{k,\nu})\rangle|\psi_k\rangle\mid^2\nonumber\\
&&\approx A_{n}\sum_{\nu,q}\mid\langle\psi_j|\mathcal{M}^{\nu}(q \cdot r)|\psi_k\rangle\mid^2\mid\langle\chi_j(Q_{j,\nu})|\chi_k(Q_{k,\nu})\rangle\mid^2\nonumber\\
&&=\mid \mathcal{M}_{jk}^{SO}+ \mathcal{M}_{jk}^{LA}\mid^2\nonumber\\
&&\quad\times A_{n}\sum_{\nu}\mid\langle\chi_j(Q_{\nu}+\Delta_{j,\nu})|\chi_k(Q_{\nu}+\Delta_{k,\nu})\rangle\mid^2,
\end{eqnarray}
where $\mathcal{M}_{jk}^{\nu}$ is the transition matrix between the intermediate state $j$ and $k$, $\Delta_{j,\nu}=\langle\psi_j|\mathcal{M}^{\nu}(q \cdot r)|\psi_j\rangle$ denotes the shift of the equilibrium position of lattice vibration induced by the electron state $\psi_j$. In general, multiphonons Raman scattering were calculated based on the complicated high-orders perturbation processes, which are replaced by the simple overlap integral of the lattice vibration in the Huang-Rhys model. Calculating the thermal average over ($A_n$) the initial states $j$ and summing over the final states $k$, Eq. (4) can be converted into
\begin{eqnarray}
&&\mid \mathcal{M}_{jk}^{SO}+\mathcal{M}_{jk}^{LA}\mid^2[\frac{N_{SO}+1}{{N_{SO}}}]^\frac{P}{2}[\frac{N_{LA}+1}{{N_{LA}}}]^\frac{1}{2}\nonumber\\
&&\times\exp[-{F_{SO}(2N_{SO}+1)}]I_P[2F_{SO}\sqrt{N_{SO}(N_{SO}+1)}]\nonumber\\
&&\times\exp[-{F_{LA}(2N_{LA}+1)}]I_1[2F_{LA}\sqrt{N_{LA}(N_{LA}+1)}],\nonumber\\
\end{eqnarray}
where $N_\nu=1/[\exp(\hbar\omega_\nu/K_BT)-1]$ is the Bose occupation function of phonon number, $F_\nu=\sum_{q}(\omega_{\nu, q}/2\hbar)\Delta_{jk,\nu}^2 $ is the Huang-Rhys factor standing for the strength of the lattice relaxation and $\Delta_{jk,\nu}=\Delta_{j,\nu}-\Delta_{k,\nu}$ describes the shift of the lattice oscillator position before and after the transition, I$_P$ and I$_1$ are the modified Bessel functions, $P=1,2,3\cdots$ denotes the number of the scattering SO phonons. Throughout this paper, we assume that the broadening factor $\Upsilon=$5 meV, the deformation potential constant $D=30$ eV for the LA phonon with the linear dispersion $\omega_{LA}$ = $c$ $q$ $(c=7.6\times10^3m/s)$. The SO phonon mode has a single frequency and the adopted parameters for the SO phonon energies and polar substrates have been listed in table I (see the supplementary material).
\begin{figure}
\centering
\includegraphics[width=3.2in,keepaspectratio]{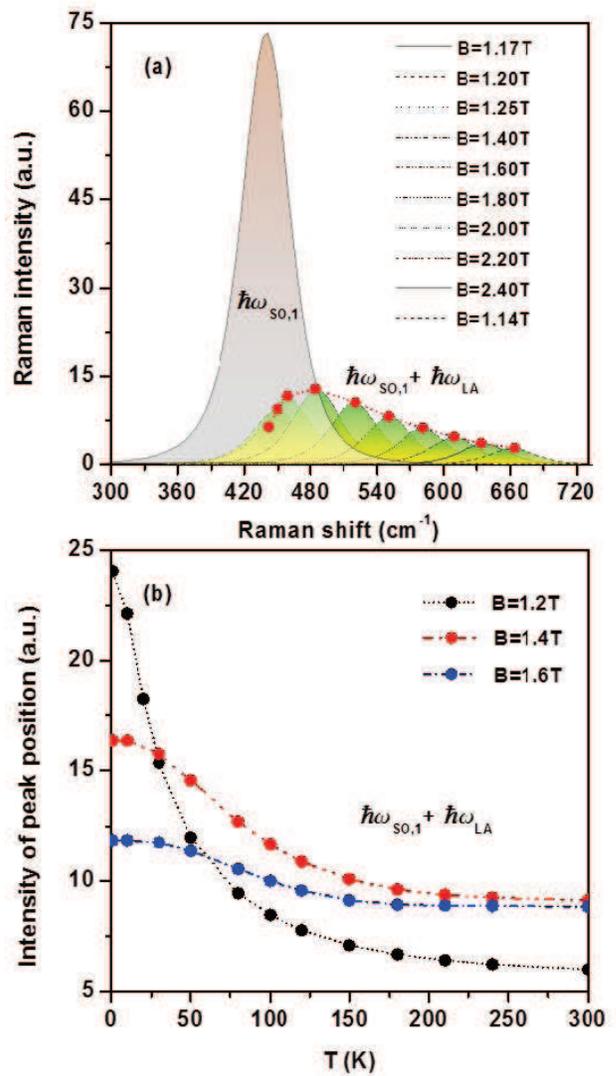}
\caption{(a) Single SO phonon ($\hbar\omega_{SO_1}$) and combined two phonons scattering (CTPS) ($\hbar\omega_{SO_1}+\hbar\omega_{LA}$) in the process of $-1\rightarrow2\rightsquigarrow0\rightarrow-1$ for different magnetic field on Al$_2$O$_3$ substrate at $T=$80 K and $z=$ 0.5 nm; (b) the temperature dependences of the CTPS ($\hbar\omega_{SO_1}+\hbar\omega_{LA}$) at different magnetic field.}
\end{figure}
\begin{figure*}
\includegraphics[width=5.8in,keepaspectratio]{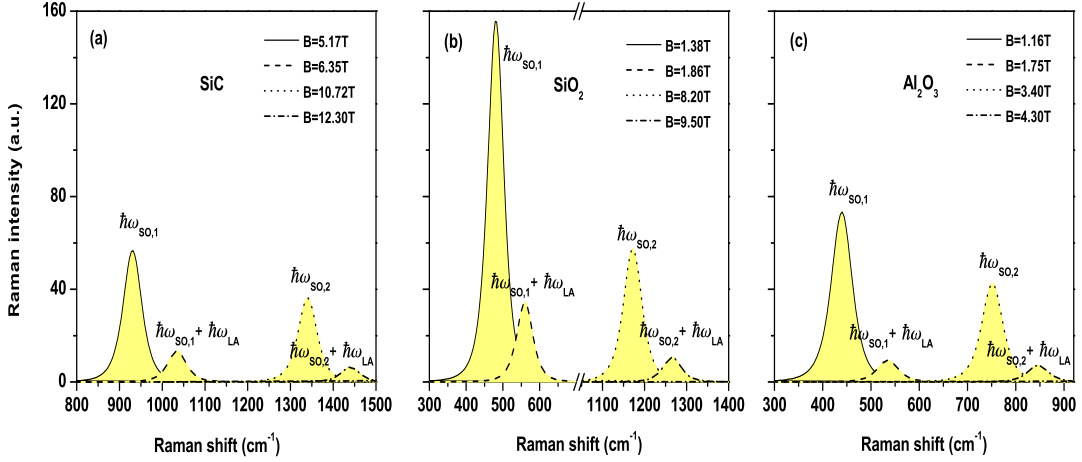}
\caption{\label{compare}Single SO phonon ($\hbar\omega_{SO_{1,2}}$) and combined two-phonons scattering ($\hbar\omega_{SO_{1,2}}+\hbar\omega_{LA}$) in the process of $-1\rightarrow2\rightsquigarrow0\rightarrow-1$ for the SiC, SiO$_2$, Al$_2$O$_3$ substrates at $T=$80 K and $z=$ 0.5 nm in (a), (b) and (c), respectively.}
\end{figure*}

Fig. 2 displays multiphonons resonance scattering in the process of $-1\rightarrow2\rightsquigarrow0\rightarrow-1$ on Al$_2$O$_3$ substrate at fixed temperature and internal distance. The absorption of incident photon and emission of scattering photon take place in the processes of $-1\rightarrow2$ and $0\rightarrow-1$, respectively, following the optical selection rule. Multiphonons scattering is in the process of $2\rightsquigarrow0$. We present the single SO phonon ($\hbar\omega_{SO_1}$ and $\hbar\omega_{SO_2}$) and the combined two-phonons scattering (CTPS) ($\hbar\omega_{SO_1}+\hbar\omega_{LA}$ and $\hbar\omega_{SO_2}+\hbar\omega_{LA}$) at specific magnetic filed. Obviously, the intensity of latter is much weaker than the former, so the CTPS may be reflected by the shoulder of the Raman spectrum of optical phonon resonance scattering in experiments\cite{wl3,wl14}. The two-SO phonons and combined three-phonons scattering are presented in Fig. 2 as the magnetic filed strong enough, but the corresponding intensity reduces drastically plotted in inset of Fig. 2. In fact, we have also examined the intrinsic LO phonon ($\hbar\omega_{LO}$) and the CTPS ($\hbar\omega_{LO}+\hbar\omega_{LA}$) and found that the intensity is reduced by several orders of magnitude than the SO phonon-assisted scattering. These multiphonons scattering could also be happened in a serials of the possible transition processes by adjusting the magnetic field, such as the process of $-2\rightarrow3\rightsquigarrow1\rightarrow-2$ plotted in supplementary materials, $-3\rightarrow4\rightsquigarrow2\rightarrow-3$ and $-2\rightarrow3\rightsquigarrow-1\rightarrow-2$. Therefore, the LQG provides a wonderful platform to identify multiphonons resonance  Raman scattering in experiments. Recently, Wendler et al proposed the flexural phonon mode is activated in low quality sample of LQG due to the defecte-assisted electron-phonon scattering\cite{wl16}. Multiphonons resonance scattering based on this kind of phonon mode should be taken into account in future.

The energy separation between LLs is modulated by the external magnetic field. The optical phonon resonance scattering doesn't occur until the energy separation matching the energy of optical phonon due to the single frequency of optical phonon is taken into account. This means that the single and multiple SO phonons scattering only take place at certain magnetic field plotted in Fig. 2. This also results in the optical phonon resonance scattering is suppressed seriously as the energy separation mismatching the energy of optical phonon. However, the combined multiple phonons scattering are allowed, such as the scattering processes of $\hbar\omega_{SO_1}+\hbar\omega_{LA}$ and $2\hbar\omega_{SO_1}+\hbar\omega_{LA}$, because of the LA phonon is included with the linear dispersion making up the mismatching. We illustrate the magnetic field and temperature dependences of the CTPS $\hbar\omega_{SO_1}+\hbar\omega_{LA}$ in Figs. 3 (a) and (b), respectively. In Fig. 3 (a), the CTPS shows two remarkable features. One is the intensity of the CTPS has a blue-shift behavior with increasing magnetic field, which attributes to the linear dispersion of LA phonon. Anther is the intensity of CTPS shows a maximum at certain magnetic field. This feature can be attributed to the interplays among three determining factors of the transition matrix element $\mathcal{M}_{jk}^{\nu}$, the strength of lattice relaxation $F_{\nu}$ and the phonon number $N_{\nu}$ in Eq. (5). With increasing magnetic field, the phonon number function $N_{LA}$ decreases sharply ($N_{SO}$ is a constant due to the single frequency), while the strength of lattice relaxation and transition matrix increase. These two opposite trends result in the appearance of the maximums for the intensity of CTPS. Fig. 3 (b) presents the intensity of CTPS as a function of the temperature at three different magnetic field. We find that the intensity decreases sharply in low temperature regime, and then vary smoothly in high temperature regime. Moreover, three curves intersect each other in low temperatures, which indicates the maximum of intensity will shift to the smaller magnetic field. This phenomena can also be traced back to the interplays between three determining factors above mentioned in Fig. 3 (a). In fact, the combined three-phonons scattering follows the similar magnetic field and temperature dependences as the CTPS.

Fig. 4 shows the influence of the polar substrate on the intensities of the single SO phonon scattering and CTPS. The SO phonons are induced by the polar substrates, so the polarizibility of substrate influences the strength of the carriers-SO phonon coupling directly, and thus the intensities of these multiphonons scattering. The polarizibility of substrate can be analyzed qualitatively by the parameter $\eta=(\kappa_{\infty}-\kappa_0)/[(\kappa_0+1)(\kappa_{\infty}+1)]$. The values of $\eta$=0.042, 0.082, 0.164 for the SiC, SiO$_2$, Al$_2$O$_3$ substrates, respectively, listed in table I (see supplemental materials). From Eqs. (3) and (5), we conclude that the transition matrix $\mathcal{M}_{jk}^{SO}$ and the modified Bessel function $I_P$ increase with the parameter $\eta$, contrasting to the exponential term due to the enhanced Huang-Rhys factor $F_{SO}$. The opposite variational trends result in the nonlinear dependence of intensity of phonon scattering on the polarizibility of substrate, similar to the magnetic field dependence of intensity discussed in Fig. 3. This is the reason why the intensity of the phonon scattering on SiO$_2$ substrate is more stronger than that of on Al$_2$O$_3$ substrate from the comparisons between Fig. 4 (b) and (c). In addition, the intensity of scattering decreases sharply with increasing the internal distance $z_0$ between graphene and substrate (see the supplementary material), because of the strength of carriers-SO coupling is decayed exponentially with this distance.

In conclusion, we theoretically study multiphonons resonance Raman scattering in LQG on different polar substrates. In particular, we propose the combined multiphonon scattering (CMS) which may be proved as the shoulders of the Raman spectrum of optical phonon resonance. We find that (i) the CMS has a blue-shift behavior with increasing magnetic field and its intensity reaches a maximum at certain magnetic field; (ii) the intensity of CMS decreases sharply in low temperature regime, and then vary smoothly in high temperature regime. Moreover, the maximum of the scattering intensity shifts to the smaller values of magnetic field at low temperature regime; (iii) the nonlinear dependences of the scattering intensity on the polarizibility of substrates are shown in the present model.

This work was supported by National Natural Science Foundation of China (No 11674241 and No 11304355).

\end{document}